\documentclass[3p, 11pt]{elsarticle}
\usepackage{graphicx,amscd,amsmath,amssymb,verbatim}
\usepackage{amsfonts,epsfig}
\usepackage{mathptmx}
\usepackage{setspace}
\usepackage{epsfig}
\usepackage{url}
\usepackage{multirow}
\usepackage{color}
\usepackage{subfigure}

\begin{document}

\journal{Elsevier}

\begin{frontmatter}

\title{Highly detailed computational study of a surface reaction model with diffusion: 
four algorithms analyzed via time-dependent and steady-state Monte Carlo simulations}

\author{Roberto da Silva$^{1}$, Henrique A. Fernandes$^{2}$}

\address{$^1$Instituto de F{\'i}sica, Universidade Federal do Rio Grande do Sul, Av. Bento Gon{\c{c}}alves, 9500 - CEP 91501-970, Porto Alegre, Rio Grande do Sul, Brazil\\
$^2$Instituto de Ci{\^e}ncias Exatas, Universidade Federal de Goi{\'a}s, Regional Jata{\'i}, BR 364, km 192, 3800 - CEP 75801-615, Jata{\'i}, Goi{\'a}s, Brazil}


\begin{abstract}

In this work, we present an extensive computational study on the Ziff-Gulari-Barshad (ZGB) model extended in order to include the spatial diffusion of oxygen atoms and carbon monoxide molecules, both adsorbed on the surface. In our approach, we consider two different protocols to implement the diffusion of the atoms/molecules and two different ways to combine the diffusion and adsorption processes resulting in four different algorithms. The influence of the diffusion on the continuous and discontinuous phase transitions of the model is analysed through two very well established methods: the time-dependent Monte Carlo simulations and the steady-state Monte Carlo simulations. We also use an optimization method based on a concept known as coefficient of determination to construct color maps and obtain the phase transitions when the parameters of the model vary. This method was proposed recently to locate nonequilibrium second-order phase transitions and has been successfully used in both systems: with defined Hamiltonian and with absorbing states. The results obtained via time-dependent Monte Carlo simulation along with the coefficient of determination are corroborated by traditional steady-state Monte Carlo simulations also performed for the four algorithms. Finally, we analyse the finite-size effects on the results, as well as, the influence of the number of runs on the reliability of our estimates. 
\end{abstract}

\end{frontmatter}


\setlength{\baselineskip}{0.7cm}

\section{Introduction}

\label{sec:introduction}

In a pioneering work back in 1986, Ziff, Gulari, and Barshad proposed a
model that mimics reactions on a catalytic surface \cite{ziff1986} in order
to describe some nonequilibrium aspects of the production of carbon dioxide $%
(CO_{2})$ molecules through the reaction of carbon monoxide $(CO)$ molecules
with oxygen $(O)$ atoms adsorbed on the surface. This stochastic model, also
known as ZGB model, has attracted much interest due to its simplicity and
rich phase diagram with continuous and discontinuous phase transitions that
separate absorbing phases from a reactive phase with sustaintable production
of $CO_{2}$ \cite{ziff1986,meakin1987,dickman1986,fischer1989}.

In addition to the scientific interest, the reason for the increasingly
study of catalytic reaction models on surfaces lies in their possible
technological applications \cite{bond1987,zhdanov1994,marro1999}. In this
respect, the ZGB model has been vastly studied nowadays and is considered a
prototype for the study on catalytic surfaces due to its inherent reaction
processes that also take place in industry which deals with oxidation of $CO$
molecules on transition-metal catalysts.

Nevertheless, some aspects of the catalytic reaction can not be explained by
this model. While some experimental works on platinum confirm the existence
of discontinuous irreversible phase transitions (IPT) in the catalytic
oxidation of $CO$ molecules \cite%
{golchet1978,matsushima1979,ehsasi1989,christmann1991,block1993}, there is
no experimental evidence of the continuous IPT. From the theoretical point
of view, the continuous phase transition has been studied by several authors
and the results support that this transition belongs to the directed
percolation (DP) universality class \cite{janssen1989,grinstein1989}.

Modified versions of the ZGB model has been proposed in order to obtain
systems with actual catalytic processes and/or to eliminate the continuous
phase transition of the original model. Some modifications include the
desorption \cite%
{fischer1989,dumont1990,albano1992,tome1993,kaukonen1989,jensen1990a,
buendia2009} or/and the diffusion \cite%
{fischer1989,ehsasi1989,kaukonen1989,jensen1990a,grandi2002} of $CO$
molecules, the inclusion of impurities on the surface \cite%
{hoenicke2000,buendia2012,buendia2013,buendia2015,hoenicke2014}, the
consideration of attractive and repulsive interactions between the adsorbed
molecules \cite{satulovsky1992}, the analysis of surfaces with different
geometries \cite{meakin1987,albano1990}, and the study of hard oxygen
boundary conditions \cite{brosilow1993}, etc. In addition, they have been
studied through several techniques, such as simulations, mean-field
theories, series analysis, etc \cite{marro1999}.

In this paper, we perform time-dependent Monte Carlo (TDMC) simulations in
order to explore the behavior of the phase transitions of the ZGB model when
both $O$ atoms and $CO$ molecules adsorbed on the catalytic surface are able
to move, i.e., when diffusion is allowed according to a protocol of
diffusion which we call DIF-I. We also consider in this work another
protocol, named as DIF-II, which was used by other authors \cite{jensen1990a}
in order to compare or even to classify the different results obtained with
these protocols. Moreover, for each protocol, we also consider two different
ways (the prescriptions) to combine the diffusion and adsorption processes
which take place on the catalytic surface. Therefore, the choice of one
prescription along with one diffusion protocol defines one algorithm. So,
when considering all possibilities of choice, we have four different
algorithms to be studied in this work. In addition, we also carry out
standard steady-state Monte Carlo (SSMC) simulations in order to support our
results.

By using TDMC simulations and an optimization method based on the concept of
the coefficient of determination \cite{trivedi2002}, we construct maps and
obtain the transitions when the parameters of the model vary. As shown in
the next section, the only two parameters considered in our study are the
rate of adsorption of $CO$ molecules and the rate of diffusion of $CO$
molecules and of $O$ atoms on the catalytic surface. This technique,
initially proposed by da Silva et. al\cite{roberto2012}, has been used in
several works such as in the study of reversible systems \cite%
{roberto2013a,roberto2013b,roberto2014,Fernandes2017} and to determine the
critical immunization probability of an epidemic model \cite{roberto2015},
as well as to obtain the continuous transition point and the upper spinodal
point of the ZGB model \cite{fernandes2016}.

Our work is organized as follows: In the next section, we present the model
and show the two prescriptions considered according to the diffusion and
adsorption processes. In Sec. \ref{sec:simulations} we describe some
finite-size scaling relations in nonequilibrium systems with absorbing
states and show the power laws obtained at the criticality as well as the
optimization method based on the coefficient of determination used to locate
the phase transitions. In this same section we also present some
considerations about the Monte Carlo method used in this work. Our main
numerical results are shown in Sec. \ref{sec:results}. Finally, the
summaries and conclusions are considered in Sec. \ref{sec:conclusions}.

\section{The Model}

\label{sec:model}

The ZGB model was devised in 1986 by R.M. Ziff, E. Gulari, and Y. Barshad 
\cite{ziff1986} in order to describe the production of carbon dioxide $%
(CO_{2})$ molecules through the reaction of carbon monoxide $(CO)$ molecules
with oxygen $(O)$ atoms on a catalytic surface. In other words, the ZGB
model is a dimer-monomer model which simulates the catalysis between the
carbon monoxide molecule and the oxygen atom \cite{ziff1986,evans1991b}. The
original model possesses three well known reactions, as follows: 
\begin{eqnarray}
CO(g)+V &\rightarrow &CO(a),  \label{eq01} \\
O_{2}(g)+2V &\rightarrow &2O(a),  \label{eq02} \\
CO(a)+O(a) &\rightarrow &CO_{2}(g)+2V,  \label{eq03}
\end{eqnarray}%
In these reactions, $V$ stands for vacant sites, and $g$ and $a$ refer,
respectively, to the gas and adsorbed phases of the atoms/molecules. As
shown in the above equations, $O_{2}$ and $CO_{2}$ molecules exist only in
the gas phase $(g)$, $O$ atoms are present only in the adsorbed phase $(a)$,
and $CO$ molecules are able to be in both phases.

In order to simulate such a model, one can think of the catalytic surface as
a regular square lattice with its sites occupied by $CO$ molecules or by $O$
atoms or even be vacant $(V)$. So, the simulation is carried out as follows 
\cite{ziff1986,marro1999,albano1996}: By following Eq. (\ref{eq01}), the $CO$
molecule in the gas phase is chosen to impinge on the surface with a rate $%
y_{CO}$. The molecule strikes the lattice in a site previously chosen at
random. If the site is vacant $(V)$, then the $CO$ molecule is adsorbed on
it. Otherwise, if the site is occupied by a $CO$ molecule or by an $O$ atom,
the trial ends, the $CO$ molecule returns to the gas phase, and a new
molecule is chosen. On the other hand, from Eq. (\ref{eq02}), if an $O_{2}$
molecule in the gas phase is chosen to hit the lattice, it does so at a rate 
$y_{O_{2}}=1-y_{CO}$. In this case, a nearest-neighbour pair of sites is
chosen at random, and, if both sites are vacant $(2V)$, the $O_{2}$ molecule
dissociates into a pair of $O$ atoms which are immediately adsorbed on the
chosen lattice sites. However, if one or both sites are occupied, the trial
ends, the $O_{2}$ molecule returns to the gas phase, and a new molecule is
chosen. As can be seen, these rates are relative ones, $y_{CO}+y_{O_{2}}=1$,
and the model, as presented in Ref. \cite{ziff1986}, has a single free
parameter: $y=y_{CO}$. Eq. (\ref{eq03}) is related to the reaction between
the $O$ atom and the $CO$ molecule, both adsorbed on the lattice.
Immediately after each adsorption event [Eqs. (\ref{eq01}) and (\ref{eq02}%
)], the nearest-neighbor sites of the adsorbed molecule/atom are checked. If
one $O-CO$ pair is found, a $CO_{2}$ molecule is formed and quits the
lattice, leaving two empty sites on it. However, if there is the possibility
of formation of two or more $O-CO$ pairs, a pair is chosen at random to quit
the lattice.

The modified version of the zgb model addressed in the present work presents
other two reactions that are related to the diffusion process of $CO$
molecules and $O$ atoms adsorbed on the lattice: 
\begin{eqnarray}
O_{i}(a)+V_{j} &\rightarrow &O_{j}(a)+V_{i}  \label{eq04} \\
CO_{i}(a)+V_{j} &\rightarrow &CO_{j}(a)+V_{i}  \label{eq05}
\end{eqnarray}%
The diffusion process may occur in two different ways which we call the
protocols DIF-I and DIF-II. For the protocol DIF-I, A site $i$ is chosen at
random. If it is occupied by an $O$ atom or by a $CO$ molecule, the
diffusion occurs with probability $p$ if at least one of its nearest
neighbor sites is vacant (site $j$). If there are two or more
nearest-neighbor vacant sites, a site $j$ is chosen at random and the
atom/molecule moves to the site $j$ leaving the site $i$ empty. It is worth
to mention that if the diffusion occurs, one must look immediately for the
formation of $CO_{2}$ molecules among the atom/molecule moved and its
nearest neighbors.

The protocol DIF-II was proposed by other authors (see, for instance, Ref. 
\cite{jensen1990a}) whose idea is similar to our proposal (DIF-I): a site $i$
is chosen at random and, if it is occupied by an $O$ atom or by a $CO$
molecule, the diffusion process may or may not occur depending on whether
the nearest neighbor is vacant or not. However, there is one important
difference: a nearest-neighbor site $j$ is chosen at random and if it is
vacant, the diffusion process occurs with a rate $p$, but if is occupied by
another atom/molecule, the trial ends and there is no diffusion no matter if
the others nearest neighbors of the site $i$ are empty or not.

The ZGB model possesses two phase transitions. The first one is a continuous
phase transition and occurs at the critical point $y_1 \cong 0.3874$ \cite%
{fernandes2016,voigt1997}. The second transition is discontinuous and occurs
at $y_2 \cong 0.5256$ \cite{ziff1992}. For $0 \leq y < y_1$ the surface
becomes irreversibly poisoned by $O$ atoms ($O-$poisoned state) and for $y_2
< y \leq 1$ the surface becomes irreversibly poisoned by $CO$ molecules ($%
CO- $poisoned state). The poisoned state, or absorbing phase, represents
states in which, once reached, the system becomes trapped and can not escape
anymore. On the other hand, for $y_1 < y < y_2$ there exists a reactive
steady state with sustainable production of $CO_2$ molecules. So, both $y_1$
and $y_2$ are irreversible phase transition (IPT) points between the
reactive and poisoned states.

In this study, we are wondering if the diffusion of the atoms/molecules
influences the first and second order phase transitions of the ZGB model. In
order to answer this question, we separate the MC simulations into two
different moments: the first moment is related to the diffusion process
given by Eqs. (\ref{eq04}) and (\ref{eq05}), and the second one deals with
the catalytic reaction given by Eqs. (\ref{eq01}), (\ref{eq02}), and (\ref%
{eq03}). In addition, we proposed two different ways of performing these
simulations:

\begin{enumerate}
\item \textit{Prescription} I: Diffusion process \textquotedblleft
XOR\textquotedblright\ adsorption process -- With a rate $p$, the diffusion
process is chosen (first moment). On the other hand, the adsorption process
occurs with a rate $1-p$, and, in this case, it is governed by the parameter 
$y$ as well as by the rules defined previously (second moment). In summary,
both the diffusion and adsorption processes are allowed but not in the same
trial.

\item \textit{Prescription }II: Diffusion process \textquotedblleft
OR\textquotedblright\ adsorption process -- The diffusion process of the
atoms/molecules occurs with a rate $p$ and then the adsorption process can
occur with a rate $y$. In summary, both process can occur in the same trial.
\end{enumerate}

So, combining the two prescriptions and the two protocols for the diffusion,
we have four algorithms which are summarized in Table \ref{Table:Algorithms2}%
.

\begin{table}[tbp] \centering%
\begin{tabular}{|l|l|l|}
\hline
\textbf{Prescription/Difusion} & DIF-I & DIF-II \\ \hline
Prescription I & Algorithm I & Algorithm III \\ 
Prescription II & Algorithm II & Algorithm IV \\ \hline
\end{tabular}%
\caption{The four algorithms considered in this study according to the
prescriptions and diffusion protocols}\label{Table:Algorithms2}%
\end{table}%

\section{Monte Carlo simulations}

\label{sec:simulations}

In this work, we study the ZGB model with diffusion of $O$ atoms and $CO$
molecules adsorbed on the catalytic surface by means of two very well
stablished simulation techniques: The time-dependent Monte Carlo (TDMC)
simulation technique, which is presented in the next subsection in some
detail along with an optimization method based on the concept of the
coefficient of determination, and the steady-state Monte Carlo (SSMC)
simulation technique, which is a traditional simulation approach and,
therefore, is presented briefly in subsection \ref{subsec:ssmc}. In all
simulations, we used periodic boundary conditions. In the study of the model
via TDMC simulations, we consider square lattices of linear size $L=80$, $%
N_{run}=5000$, and $t_{MC}=300$. Here, $N_{run}$ stands for the number of
independent time series which are considered to obtain the averages which,
in turn, compose the curve $\rho (t)\times t$ and that results in one value
of $r$, and $t_{MC}$ is the number of Monte Carlo (MC) steps used in each
time series. To obtain our estimates, the first $t_{mic}=100$ MC steps are
discarded. The reasons for choosing these values of $L$ and $N_{run}$ in our
simulations are discussed at the end of the Section \ref{sec:results}. Our
results show that they are enough to garantee that our results are reliable
and robust. On the other hand, for the SSMC simulations we consider $L=80$
and $t_{\max }=10^{5}$ MC steps to average the densities. In order to make
sure that the steady state has been reached before taking the averages, we
discarded the first $t_{disc}=10^{3}$ MC steps.

\subsection{Short-time dynamics and time-dependent Monte Carlo simulations
in models without defined Hamiltonian}

\label{subsec:tdmc}

In order to perform the numerical simulations, we take into consideration
that, for systems belonging to the DP universality class, the finite-size
scaling near criticality can be described by: 
\begin{equation}
\left\langle \rho (t)\right\rangle \sim t^{-\beta /\nu _{\parallel
}}f((y-y_{c})t^{1/\nu _{\parallel }},t^{d/z}L^{-d},\rho _{0}t^{\beta /\nu
_{\parallel }+\theta }),  \label{eqfss}
\end{equation}
where $\rho$ is the density that may be of $CO$ molecules ($\rho_{CO}$) or
of vacant sites ($\rho_V$). The $\left\langle \cdots \right\rangle $ stands
for the average on different evolutions of the system, $d$ is the dimension
of the system (for the ZGB model, $d=2$), $L$ is the linear size of a
regular square lattice, and $t$ is the time. The indexes $z=\nu _{\parallel
}/\nu _{\perp }$ and $\theta =\frac{d}{z}-\frac{2\beta }{\nu _{\parallel }}$
are dynamic critical exponents, and $\beta $, $\nu _{\parallel }$, and $\nu
_{\perp }$ are static ones. Here, $y-y_{c}$ denotes the distance of a point $%
y$ to the critical point, $y_{c}$, which governs the algebraic behaviours of
the two independent correlation lengths: the spatial one which behaves as $%
\xi _{\perp }\sim (y-y_{c})^{-\nu _{\perp}}$ and the temporal one, $\xi
_{\parallel }\sim (y-y_{c})^{-\nu _{\parallel}}$.

As stated above, this model possesses two different order parameters, $\rho
_{CO}$ and ($\rho _{V}$) and so, we can use any of them to obtain our
results. When considering TDMC simulations, most of our results are obtained
by considering the density of vacant $V$ sites as the order parameter since
it is more stable than those of the density of $CO$ molecules near the
criticality (when the fluctuations become important), thereby producing
better results. However, these fluctuations do not prevent us from studying
the model with the density of $CO$ molecules mainly when considering the
region of the discontinuous phase transition since, in that case, the
fluctuations are much less prominent than for the continuous one. This
density is given by 
\begin{equation*}
\rho (t)=\frac{1}{L^{d}}\sum_{j=1}^{L^{d}}s_{j},
\end{equation*}%
where $s_{j}=1$ when the site $j$ is empty (for $\rho _{V}$) or is occupied
by a $CO$ molecule (for $\rho _{CO}$). Otherwise, it is equal to zero.

As shown in Ref. \cite{fernandes2016}, the dynamic and static critical
exponents of the model can be obtained by using the Eq. (\ref{eqfss}) and
performing TDMC simulations with two different initial conditions $\rho
(0)=\rho_{0}$: (i) the lattice is completely empty, i.e. there exist only
vacant sites ($\rho_{0}=0$), and (ii) the lattice is completely filled with $%
O$ atoms but a random site which remains empty ($\rho _{0}=1/L^{2}$). When
considering the first condition, one obtains 
\begin{equation}
\left\langle \rho \right\rangle (t)\sim t^{\lambda },  \label{eq_pl1}
\end{equation}
where $\lambda =-\beta /\nu_{\parallel }$, and the second condition yields 
\begin{equation}
\left\langle \rho \right\rangle (t)\sim \rho _{0}t^{\theta }=\rho
_{0}t^{\left( \frac{d}{z}-2\frac{\beta }{\nu _{\parallel }}\right) }.
\label{eq_pl2}
\end{equation}

The numerical computing of the exponents $z$, $\beta $, and $\nu _{\parallel
}$, follows a straightforward procedure: First, by considering these two
different initial conditions, it is possible to obtain the exponent $z$ in a
very simple way, leading to the following power law $F_{2}(t)=\left\langle
\rho \right\rangle _{\rho _{0}=1/L}(t)/\left\langle \rho \right\rangle
_{\rho _{0}=1}^{2}(t)\sim t^{d/z}$. In addition to the analysis of these
power laws in Ref. \cite{fernandes2016}, this idea has been applied
successfully in a large number of spin systems: for example, the Ising
model, the $q=3$ and $q=4$ Potts models \cite{Silva20021}, the Heisenberg
model \cite{Fernandes2006c} and even for models based on the generalized
Tsallis statistics \cite{roberto2012}. It was firstly used in systems
without defined Hamiltonian in 2004, as can be seen in Ref. \cite%
{Roberto2004}. Second, in order to compute $\nu _{\parallel }$, another
power law is obtained when taking into account the following derivative \cite%
{Grassberger1996}: $D(t)=\frac{\partial \ln \left\langle \rho \right\rangle 
}{\partial y}\bigg\vert_{y=y_{c}}$ which is numerically represented by $D(t)=%
\frac{1}{2\delta }\ln \left( \frac{\left\langle \rho \right\rangle
(y_{c}+\delta )}{\left\langle \rho \right\rangle (y_{c}-\delta )}\right) ,$%
where $\delta $ is a tiny perturbation needed to move the system slightly
off the criticality, yelding a power law decay that only depends on $\nu
_{\parallel }$, i.e, $D(t)=t^{\frac{1}{\nu _{\parallel }}}$.

However, to determine these exponents through this set of power laws, the
critical point must be accurately determined. But can we use TDMC
simulations to perform this task? The answer is yes, we can! Actually, this
is our goal, i.e., we are concerned with the influence of the diffusion on
the continuous and discontinuous irreversible phase transitions of the ZGB
model. So, instead of focusing on the estimate of the critical exponents, we
decided to take into account a very simple and efficient optimization method
to obtain an overview of the phase transitions when the rate of adsorption
of $CO$ molecules $(y)$ on the surface and the rate of diffusion $p$ vary
from zero to one, i.e., $0\leq y\leq 1$ and $0\leq p\leq 1$. This
optimization method considers the robustness of the power laws and is based
on a quantity known as coefficient of determination. It is given by 
\begin{equation}
r=\frac{\sum\limits_{t=t_{mic}}^{t_{MC}}(\overline{\ln \left\langle \rho
\right\rangle }-a-b\ln t)^{2}}{\sum\limits_{t=t_{mic}}^{t_{MC}}(\overline{%
\ln \left\langle \rho \right\rangle }-\ln \left\langle \rho \right\rangle
(t))^{2}},  \label{determination_coefficient}
\end{equation}%
where $\rho =\rho (t)$ is obtained for each pair ($y,p$), $t_{MC}$ is the
number of Monte Carlo (MC) steps, $a$ and $b$ are, respectively, the
intercept and the slope of a linear function, and $\overline{\ln
\left\langle \rho \right\rangle }=\frac{1}{(t_{MC}-t_{mic})}%
\sum\nolimits_{t=t_{mic}}^{t_{MC}}\ln \left\langle \rho \right\rangle (t)$,
with $t_{mic}$ the number of MC steps that are discarded in our TDMC
simulations. Here, differently from $\overline{O}=$ $\frac{1}{%
(t_{MC}-t_{mic})}\sum\limits_{t=t_{mic}}^{t_{MC}}O(t)$ which denotes an
average over the MC steps, $\left\langle O\right\rangle
(t)=(1/N_{run})\sum\limits_{i=1}^{N_{run}}O_{i}(t)$ denotes an average over
the different runs, i.e., an average over different time evolutions
considering different set of random numbers.

This coefficient ranges from 0 to 1 and measures the ratio: (expected
variation)/(total variation), so that the bigger the $r$, the better the
linear fit of the data in log-scale. When the system is out of criticality,
there is no power law and $r \simeq 0$. However, at criticality ($y=y_c$ and 
$p=p_c$) or close to it, it is expected that $\rho(t)$ possesses a power law
behavior given, for instance, by Eq. (\ref{eq_pl1}) and $r$ approaches to 1.

Therefore, we perform TDMC simulations for each pair $y_{i}=i\Delta $ and $%
p_{j}=j\Delta $, with $i,j=0,...,n$ and $n=1/\Delta $, in order to obtain
color maps for the coefficient of determination as function of each pair $%
(y,p)$. Such maps are able to show the existence (or not) of phase
transitions when diffusion of $O$ atoms and $CO$ molecules are allowed. The
pair of critical values $(y_{c},p_{c}$ corresponds to $%
(y^{(opt)},p^{(opt)})=\arg \max_{y\in \lbrack 0,1],p\in \lbrack 0,1]}\{r\}$.
A similar procedure has been used by the authors to study the effects of
mobility on diluted lattices in an epidemic model \cite{roberto2015}. In
that work, they showed that mobility influences the immunization rates in
such a way that the greater the mobility, the bigger the immunization rate.

It is important to mention that this technique can also be extended to study
the weak first-order phase transitions \cite{fernandes2016}, since these
transitions possess long correlation lenghts and small discontinuities and
therefore behave similarly to second-order phase transitions \cite%
{Schulke2001,Albano2001}. It has been conjectured that near a weak
first-order transition there exist two pseudo-critical points: one point is
just below (lower) the first-order phase transition point, and the other one
is just above (upper) it. These pseudo-critical points are known as spinodal
points \cite{Schulke2001,Albano2001}.

\subsection{Steady-state Monte Carlo simulations}

\label{subsec:ssmc}

In order to support our TDMC simulations, we also perform SSMC simulations
to obtain the density of $CO$ and $CO_{2}$ molecules, $O$ atoms, and vacant
sites $V$. In these simulations, the first $t_{disc}$ MC steps are discarded
to make sure the steady state has been reached. After that, the averages of
the densities are taken over $t_{\max }-t_{disc}$ simulation
steps.Therefore, the initial condition of the lattice is not as important as
in TDMC simulations. So, we consider all lattice sites empty at the
beginning of the SSMC simulations

To obtain the phase diagram of the model for the different densities, we fix
a value of $p$ and sweep $y$ from 0 to 1. In such a wah, we are able to
obtain well defined values of $y$ for the continuous and discontinuous phase
transitions for a given value of $p$.

\section{Results}

\label{sec:results}

In this work, we perform nonequilibrium MC simulations in order to obtain
the phase transitions of the ZGB model when the diffusion of $O$ atoms and $%
CO$ molecules adsorbed on the lattice is allowed. The main results of our
simulations are presented by means of color maps of the coefficient of
determination $r$ given by Eq. (\ref{determination_coefficient}) as function
of $y$ and $p$. As stressed at the end of Sec. \ref{sec:model}, we proposed
two different prescriptions: the first one considers that only one process
(adsorption or diffusion) can occur in a given trial (algorithms I and III),
and the second one in which both adsorption and diffusion processes are able
to happen simultaneously (algorithms II and IV).

Our results are organized as follows: In the first two subsections, we study
the diffusion process (DIF-I) with the algorithms I and II. These
subsections present our main results since they deal with a different
approach to explore the ZGB model with diffusion of the adsorbed
atoms/molecules. We first perform simulations in order to look into the
entire space of the parameters $y$ and $p$, i.e., $0\leq y\leq 1$ and $0\leq
p\leq 1$, and obtain an overview of the phase transitions of the model by
building color maps. We also present in Subsec. \ref{subsec1} the results
obtained through SSMC simulations performed to give support to our
nonequilibrium simulations.

In the last subsection, we complete our study by including the results
obtained through the protocol DIF-II and the algorithms III and IV. The
results obtained for the algorithm III contemplate the estimates already
found in the literature and the algorithm IV presents other possible
variation of the model. So, these four algorithms show a complete
classification of the problem which is, in this work, solved by means of two
different MC simulation techniques. It is important to mention that the
results from the literature was only studied with standard SSMC simulations
and only for algorithm III to the best of our knowledge. In that same
subsection we also discuss the possible finite size effects and the number
of runs used to perform the simulations.

\subsection{Algorithm I : Protocol DIF-I with \textquotedblleft
XOR\textquotedblright\ adsorption process}

\label{subsec1}

\begin{figure}[tbh]
\begin{center}
\includegraphics[width=0.83\columnwidth]{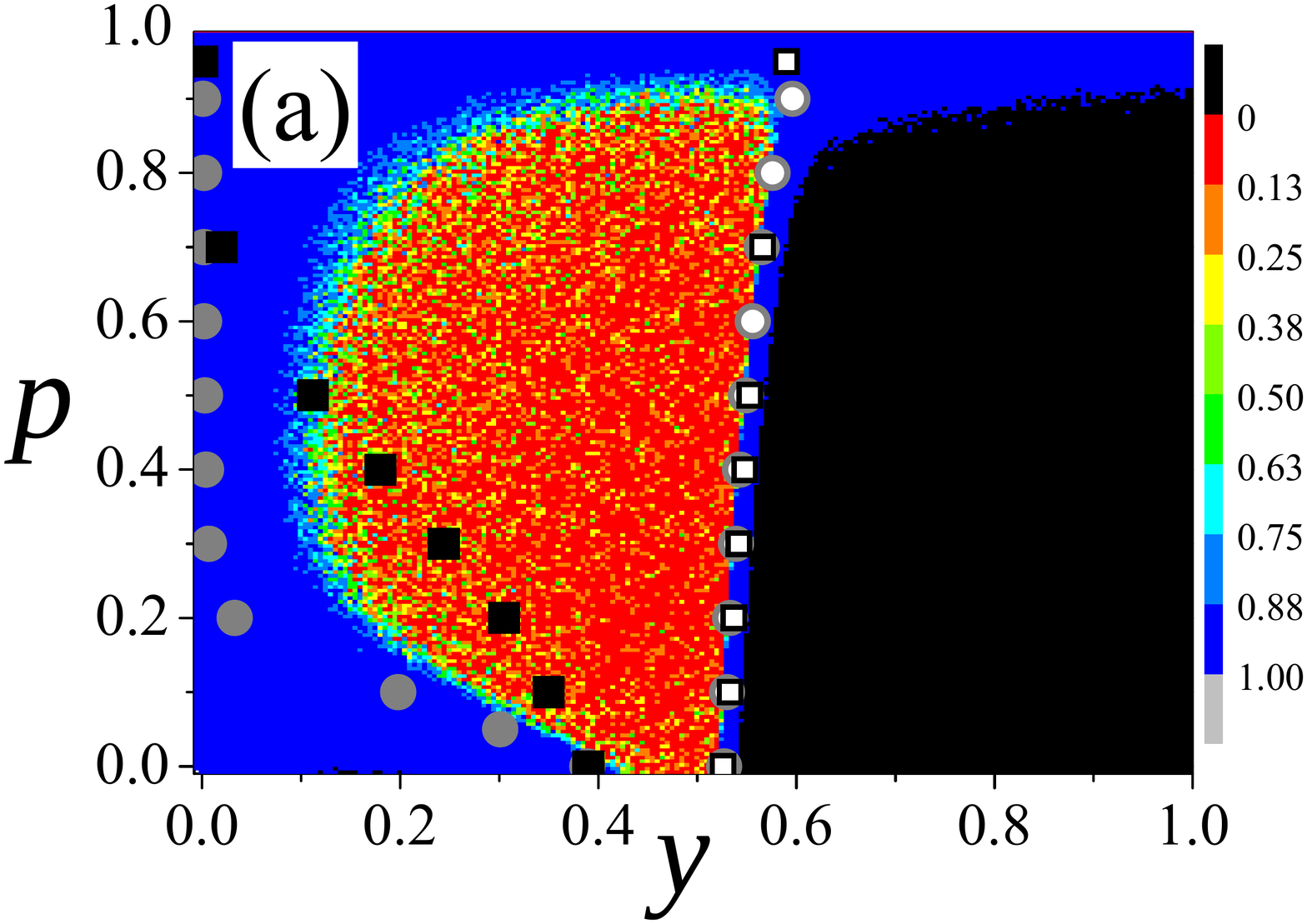} 
\includegraphics[width=0.83
\columnwidth]{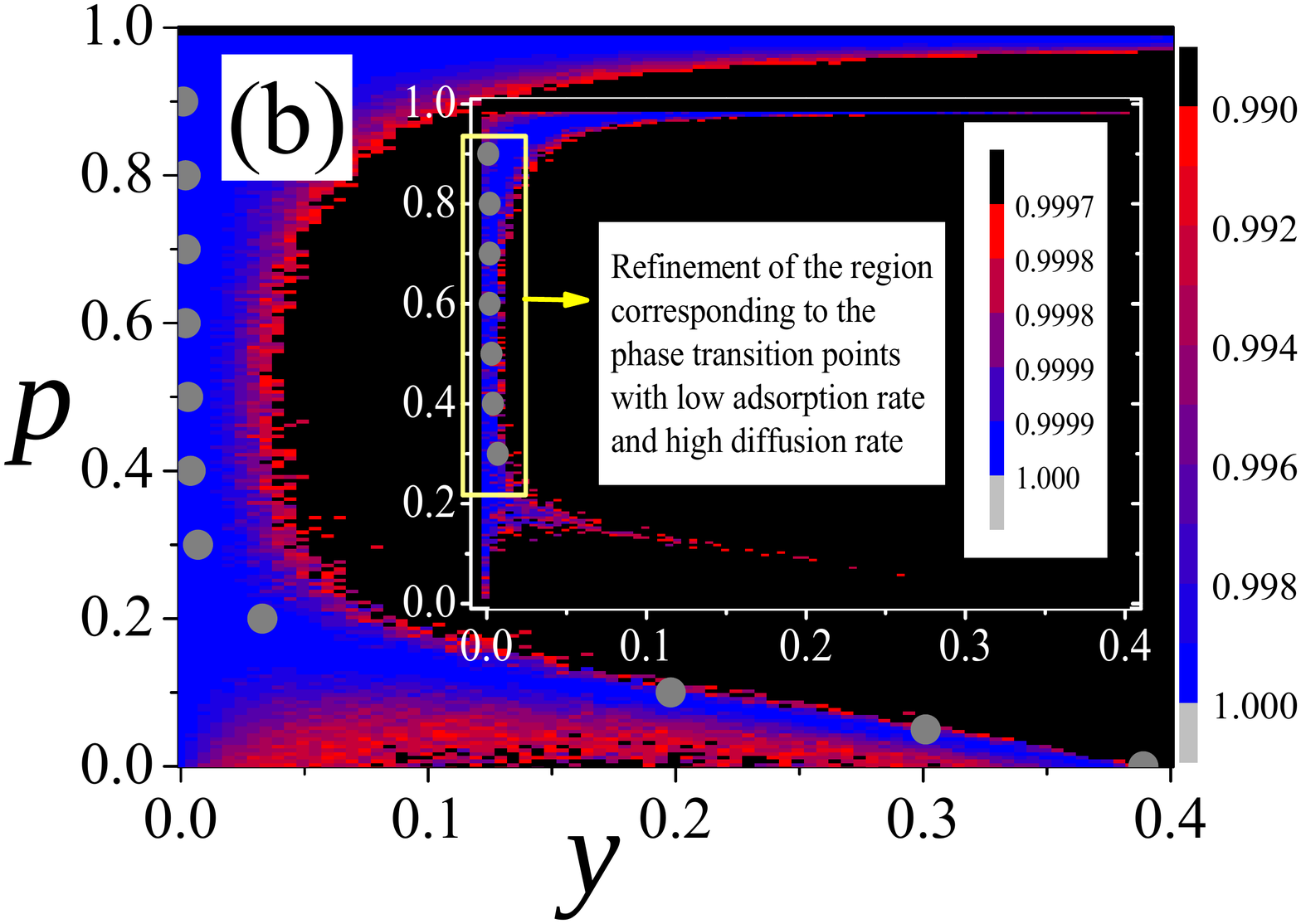}
\end{center}
\caption{(a) Coefficient of determination $r$ as function of $y$ and $p$
using the algorithm I and the TDMC simulations. The filled and empty black
squares correspond, respectively, to the continuous and discontinuous phase
transitions predicted in Ref. \protect\cite{jensen1990a}. In this same
figure, we also show the results of the SSMC simulations for the algorithm
I: The filled and empty filled gray circles correspond, respectively, to the
continuous and discontinuous transition points. (b) Refinement of region
with low adsorption rate and high diffusion rate.}
\label{Fig:space1}
\end{figure}

In Fig. \ref{Fig:space1} (a) we show the color map which represents the
possible regions of phase transitions of the ZGB model with diffusion for
the algorithm I. Each point corresponds to the coefficient of determination $%
r$ associated to power law for the density of empty sites obtained for the
pair of parameters $(y,p)$, with $0\leq y\leq 1$ and $0\leq p\leq 1$. We
change both parameters using a resolution $\Delta y=0.005$ and $\Delta
p=0.005$, resulting in more than 40000 simulations. 

As can be seen, there are two regions that deserve more attention, i.e.,
regions in which the points $(y,p)$ possess coefficient of determination
close to 1 (the blue regions in the figure). The first one is around the
continuous phase transition point of the original ZGB model, $y\cong 0.3875$%
, with $p=0$. The second one is close to $y\cong 0.525$, which is the
discontinuous phase transition point of the original ZGB model, with $p$
ranging from 0 to 1. We can observe two very different behaviors: the
discontinuous phase transition of the model is not influenced by the
diffusion of the adsorbed atoms/molecules but the continuous one is strongly
affected.

The filled and empty black squares show, respectively, the continuous and
discontinuous points estimated in Ref. \cite{jensen1990a} by using standard
SSMC simulations. Even though the protocols and algorithms are different,
one can be seen that the discontinuous phase transition points perfectly
agree with our vertical discontinuous phase transition line (blue line)
obtained through TDMC simulations. However, the continuous phase transition
points obtained in Ref. \cite{jensen1990a} present a huge difference from
our estimates since they cross the red/orange region with small coefficient
of determination, $r\approx 0-0.25 $. However, it is important to mention
that the algorithm I does not correspond to the method employed in Ref. \cite%
{jensen1990a} which is defined by the protocol II (DIF-II) and algorithm III
that is considered in the Subsec. \ref{subsec3}.

\begin{figure}[tbh]
\begin{center}
\includegraphics[width=1.0\columnwidth]{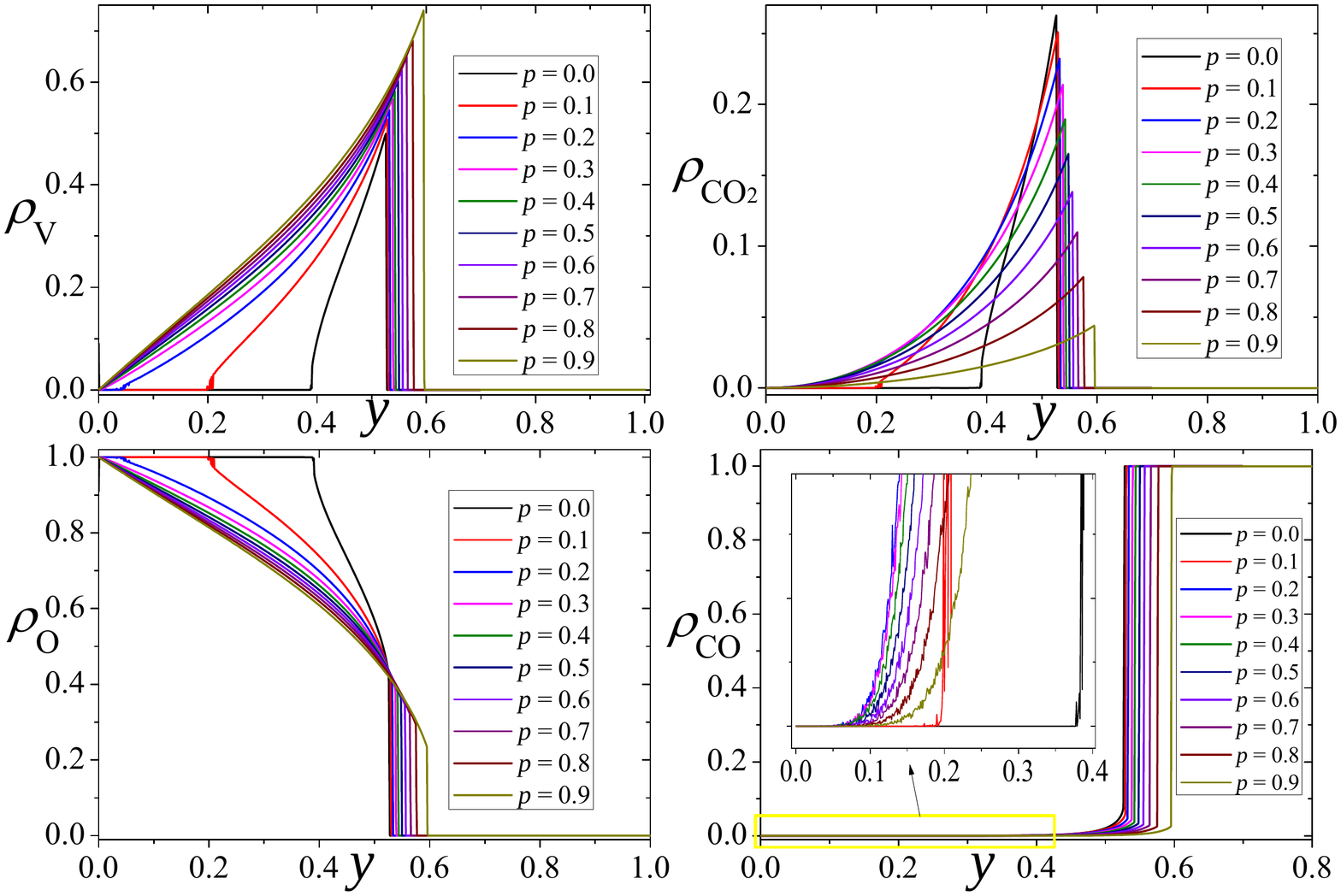}
\end{center}
\par
.
\caption{Density of empty sites $V$, $CO_{2}$ molecules, $O$ atoms, and $CO$
molecules as function of $y$, for different values of $p$, obtained through
SSMC simulations for the algorithm I. We can observe the continuous and
discontinuous phase transitions in all of these curves. In the last plot,
for density of $CO$, the inset plot shows a zoom of the continuous phase
transition region $0<y<0.4$}
\label{Fig:density_by_SSMC}
\end{figure}

In this same figure, we also show the estimates obtained through SSMC
simulations (see Fig. \ref{Fig:density_by_SSMC}) represented by filled and
empty gray circles, which refer to some continuous and discontinuous phase
transitions points, respectively, as can be seen in Table \ref{Table:ssmc2}.

\begin{table}[tbp] \centering%
\begin{tabular}{l|cccc|l|cccc|l}
\hline\hline
& \multicolumn{5}{|c|}{\textbf{Discontinuous phase transition}} & 
\multicolumn{5}{|c}{\textbf{Continuous phase transition}} \\ \hline
${\small p}$ & \multicolumn{4}{|c|}{\small Algorithm} &  & 
\multicolumn{4}{|c|}{\small Algorithm} &  \\ \hline
& {\small I} & {\small II} & {\small III} & {\small IV} & {\small Ref. \cite%
{jensen1990a}} & {\small I} & {\small II} & {\small III} & {\small IV} & 
{\small Ref. \cite{jensen1990a}} \\ \hline
{\small 0.00} & \multicolumn{1}{|l}{\small 0.527(2)} & \multicolumn{1}{l}%
{\small 0.527(1)} & \multicolumn{1}{l}{\small 0.527(1)} & \multicolumn{1}{l|}%
{\small 0.527(2)} & {\small 0.526(1)} & \multicolumn{1}{|l}{\small 0.389(1)}
& \multicolumn{1}{l}{\small 0.389(1)} & \multicolumn{1}{l}{\small 0.389(1)}
& \multicolumn{1}{l|}{\small 0.389(2)} & {\small 0.390(1)} \\ 
{\small 0.10} & \multicolumn{1}{|l}{\small 0.530(2)} & \multicolumn{1}{l}%
{\small 0.525(2)} & \multicolumn{1}{l}{\small 0.532(1)} & \multicolumn{1}{l|}%
{\small 0.530(2)} & {\small 0.533(1)} & \multicolumn{1}{|l}{\small 0.301(1)}
& \multicolumn{1}{l}{\small 0.219(1)} & \multicolumn{1}{l}{\small 0.349(1)}
& \multicolumn{1}{l|}{\small 0.352(2)} & {\small 0.350(1)} \\ 
{\small 0.20} & \multicolumn{1}{|l}{\small 0.533(2)} & \multicolumn{1}{l}%
{\small 0.522(2)} & \multicolumn{1}{l}{\small 0.537(2)} & \multicolumn{1}{l|}%
{\small 0.531(3)} & {\small 0.537(1)} & \multicolumn{1}{|l}{\small 0.198(2)}
& \multicolumn{1}{l}{\small 0.112(1)} & \multicolumn{1}{l}{\small 0.302(1)}
& \multicolumn{1}{l|}{\small 0.321(2)} & {\small 0.305(1)} \\ 
{\small 0.30} & \multicolumn{1}{|l}{\small 0.539(3)} & \multicolumn{1}{l}%
{\small 0.521(2)} & \multicolumn{1}{l}{\small 0.542(2)} & \multicolumn{1}{l|}%
{\small 0.532(3)} & {\small 0.542(1)} & \multicolumn{1}{|l}{\small 0.033(2)}
& \multicolumn{1}{l}{\small 0.059(2)} & \multicolumn{1}{l}{\small 0.249(2)}
& \multicolumn{1}{l|}{\small 0.291(3)} & {\small 0.244(1)} \\ 
{\small 0.40} & \multicolumn{1}{|l}{\small 0.543(3)} & \multicolumn{1}{l}%
{\small 0.519(2)} & \multicolumn{1}{l}{\small 0.547(2)} & \multicolumn{1}{l|}%
{\small 0.533(3)} & {\small 0.548(1)} & \multicolumn{1}{|l}{\small 0.007(3)}
& \multicolumn{1}{l}{\small 0.034(2)} & \multicolumn{1}{l}{\small 0.185(2)}
& \multicolumn{1}{l|}{\small 0.263(3)} & {\small 0.180(1)} \\ 
{\small 0.50} & \multicolumn{1}{|l}{\small 0.549(3)} & \multicolumn{1}{l}%
{\small 0.517(3)} & \multicolumn{1}{l}{\small 0.552(2)} & \multicolumn{1}{l|}%
{\small 0.533(3)} & {\small 0.553(2)} & \multicolumn{1}{|l}{\small 0.004(3)}
& \multicolumn{1}{l}{\small 0.025(3)} & \multicolumn{1}{l}{\small 0.119(2)}
& \multicolumn{1}{l|}{\small 0.236(3)} & {\small 0.112(2)} \\ 
{\small 0.60} & \multicolumn{1}{|l}{\small 0.556(4)} & \multicolumn{1}{l}%
{\small 0.514(3)} & \multicolumn{1}{l}{\small 0.558(3)} & \multicolumn{1}{l|}%
{\small 0.534(4)} & {\small -} & \multicolumn{1}{|l}{\small 0.003(3)} & 
\multicolumn{1}{l}{\small 0.022(4)} & \multicolumn{1}{l}{\small 0.056(3)} & 
\multicolumn{1}{l|}{\small 0.204(4)} & {\small -} \\ 
{\small 0.70} & {\small 0.565(4)} & {\small 0.512(4)} & {\small 0.567(3)} & 
{\small 0.534(4)} & {\small 0.566(5)} & {\small 0.002(4)} & {\small 0.017(4)}
& {\small 0.018(3)} & {\small 0.185(4)} & $\approx ${\small \ 0.02} \\ 
{\small 0.80} & {\small 0.576(5)} & {\small 0.511(4)} & {\small 0.578(5)} & 
{\small 0.534(5)} & {\small -} & {\small 0.002(5)} & {\small 0.015(5)} & 
{\small 0.006(4)} & {\small 0.164(6)} & {\small -} \\ 
{\small 0.90} & {\small 0.596(6)} & {\small 0.508(5)} & {\small 0.595(7)} & 
{\small 0.534(6)} & {\small -} & {\small 0.001(7)} & {\small 0.011(6)} & 
{\small 0.003(6)} & {\small 0.142(8)} & {\small -} \\ 
{\small 0.95} & {\small -} & {\small -} & {\small -} & {\small -} & {\small %
0.59(1)} & {\small -} & {\small -} & {\small -} & {\small -} & $\approx $%
{\small \ 0} \\ \hline\hline
\end{tabular}%
\caption{Phase transition points for the four algorithms obtained through
SSMC simulations.}\label{Table:ssmc2}%
\end{table}%

As can be seen, the discontinuous phase transition line is very robust since
all three results, obtained with different techniques and different
algorithms, are in absolut agreement with each other.

Figure \ref{Fig:space1} (b) shows a refinement of the coefficient of
determination for continuous phase transition region. This refinement
process is a straightforward procedure: we consider only values of $r$
ranging from 0.990 to 1 to construct the color map. In this case, one can
observe an interesting narrowing of the blue regions which remain containing
the filled gray circles. A further refinement of $r$ ($0.9997\leq r\leq 1$)
is showed in the inset plot of Fig. \ref{Fig:space1} (b) and confirm that
our results obtained via TDMC simulations are in complete agreement with
those ones obtained through SSMC simulations.

Figure \ref{Fig:density_by_SSMC} shows the density of empty sites $(V)$, $%
CO_2$ and $CO$ molecules, and $O$ atoms as function of $y$, for different
values of $p$. We can clearly observe the continuous and discontinuous phase
transition points when $y$ and $p$ vary. These points are, respectively, the
filled and empty gray circles already presented in Fig. \ref{Fig:space1}.

Just for completing the study of the algorithm I, we wonder about the
existence of correlation between the negativity/positivity of the exponent $%
\lambda $ and the region of maximum coefficient of determination. It is
important to mention that this is only an empirical analysis. In Fig. \ref%
{Fig:Exponent_algorithm_I}, we show the color map of the exponent $\lambda$
(Eq. \ref{eq_pl1}) as function of $y$ and $p$ for the region of the
continuous phase transition along with our results obtained by SSMC
simulations. The figure suggests that $\lambda$ lies between -1.10 and -0.7.

\begin{figure}[tbh]
\begin{center}
\includegraphics[width=0.8\columnwidth]{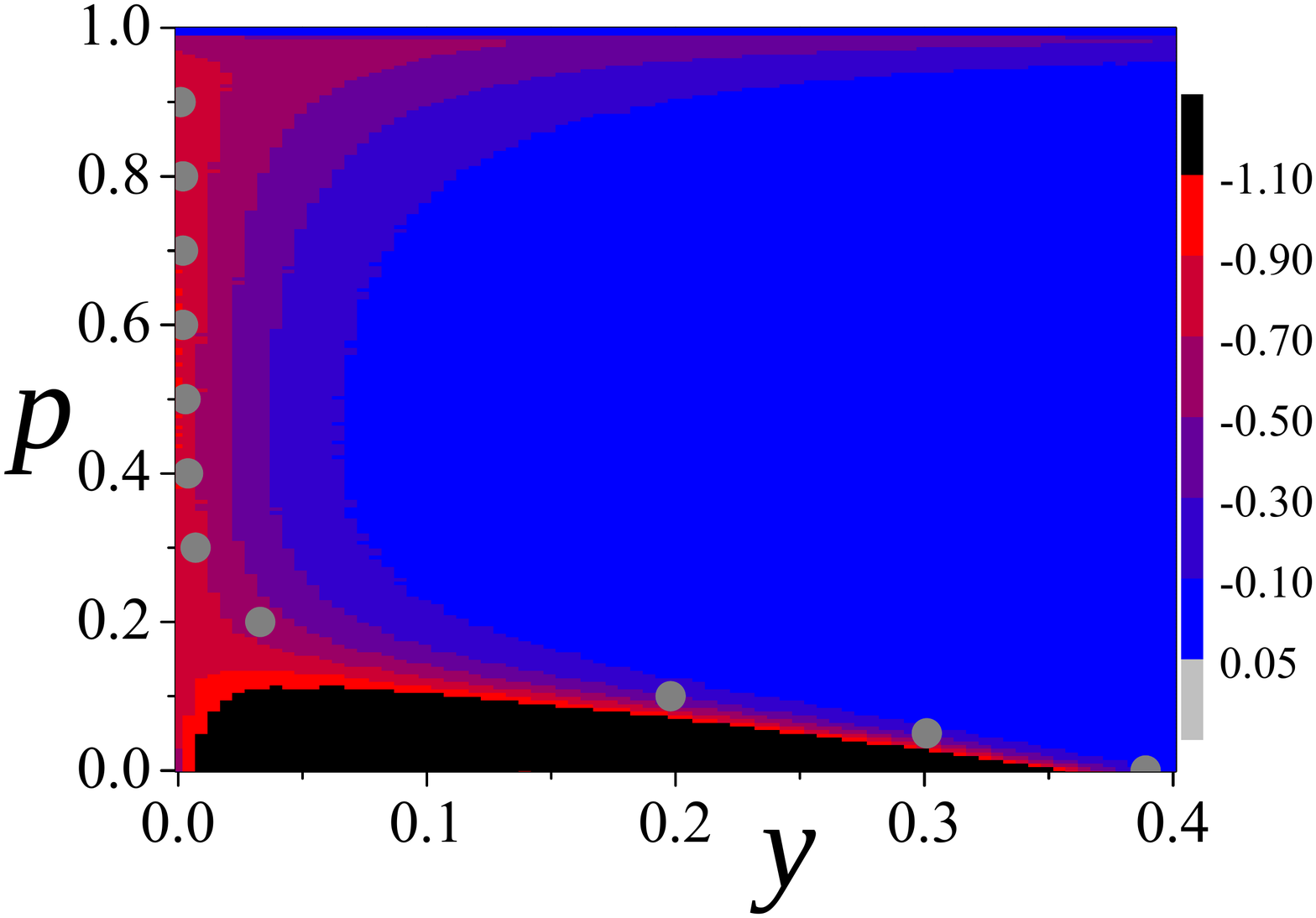}
\end{center}
\caption{Exponent $\protect\lambda $ as function of $y$ and $p$ obtained
through TDMC simulations for the algorithm I in the region of the continuous
phase transition.}
\label{Fig:Exponent_algorithm_I}
\end{figure}

As shown above, in our study, we are able to localize both continuous and
discontinuous phase transitions of the model by means of TDMC simulations
along with the coefficient of determination approach. The combination of
this two techniques has been proven to be very efficient in determining the
second order phase transition points of several models with and without
defined Hamiltonians, and with absorbing phases. At the very beginning, it
was not expected that TDMC simulations were able to estimate first order
phase transitions. However, it has been shown that it is possible to
determine, at least, weak first order phase transition points \cite%
{Schulke2001} through this technique. Recently, some works have shown the
efficiency of this combination in determination the weak first order phase
transitions of systems with and without defined Hamiltonians \cite%
{roberto2014,fernandes2016}.

From now on, we mantain the convention adopted until now for the figures,
i.e., gray circles refer to the results obtained in our SSMC simulations and
black squares denote the results taken from literature, more precisely from
the Ref. \cite{jensen1990a}. The continuous phase transition points are
represented by filled circles or squares and the discontinuous ones are
represented by empty circles or squares.

\begin{figure}[tbh]
\begin{center}
\includegraphics[width=1.0\columnwidth]{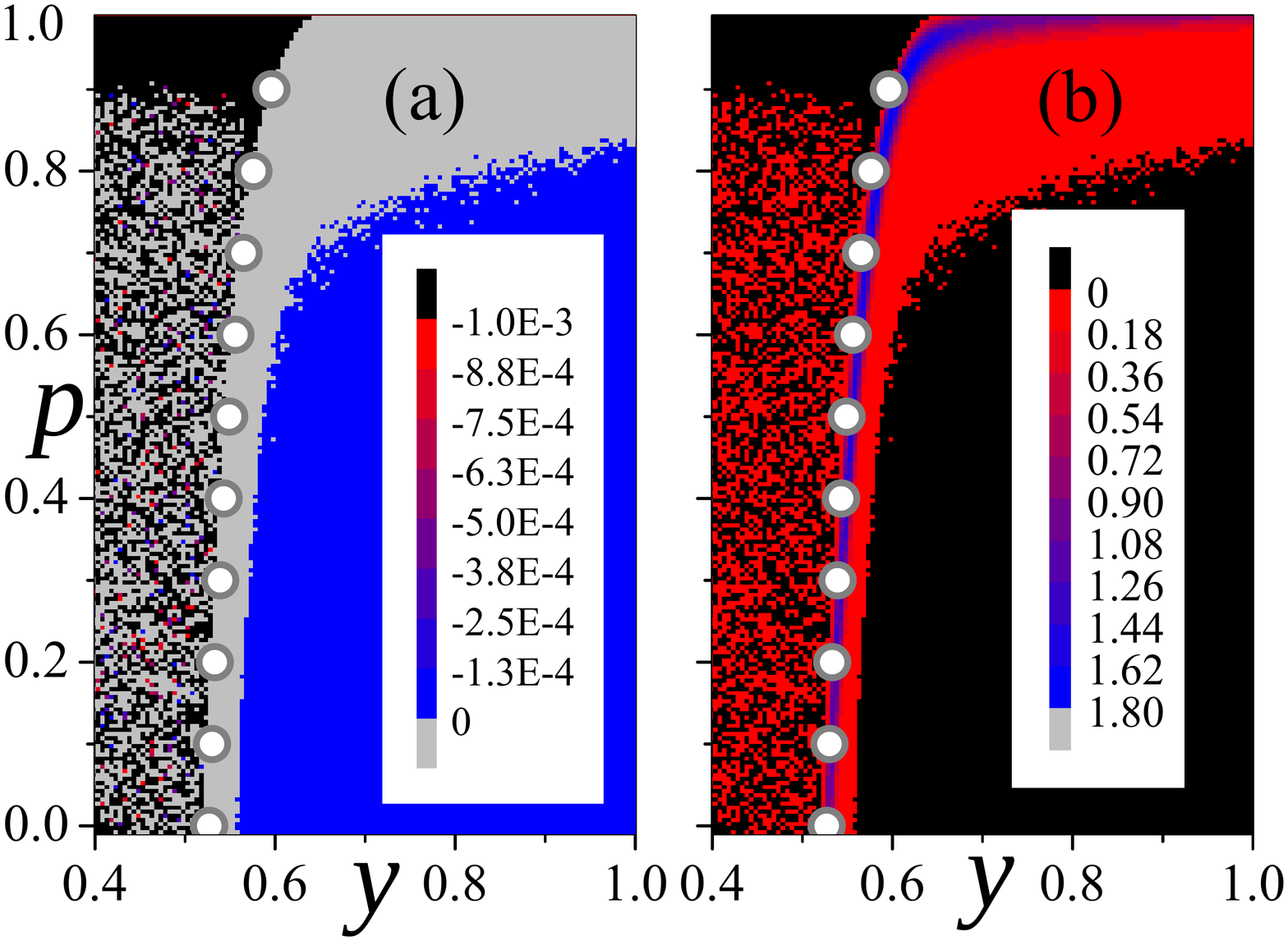}
\end{center}
\caption{(a) Exponents $\protect\lambda $ as function of $y$ and $p$
obtained through TDMC simulations for the algorithm I in the region of the
discontinuous phase transition points for $\protect\rho _{CO}$ with the
refinement $\protect\lambda \leq 0$. (b) The same plot with the refinement $%
\protect\lambda >0$.}
\label{Fig:Exponent_extra}
\end{figure}

Now, we consider the other order parameter of the model, the density of $CO$
molecules ($\rho _{CO}$), just for looking into the behavior of the exponent 
$\lambda $, i.e., its sign and absolute value, for the region of the
discontinuous phase transition. The reason for choosing this order parameter
in this analysis is that, as happen to the density of vacant sites $\rho
_{V} $, its time evolution does not present huge fluctuations in this
region. In addition, the region of interest is narrower and better defined
for $\rho _{CO}$. Figure \ref{Fig:Exponent_extra} (a)) shows the color map
of the exponent $\lambda $ as function of $y$ and $p$ when considering $\rho
_{CO}$ as the order parameter, and also the refinement $\lambda \leq 0$. We
can observe from this plot that negative exponents are discarded for the
discontinuous phase transition points. 

Figure \ref{Fig:Exponent_extra} (b) refines the region for $\lambda >0$. We
can observe that when the larger the magnitude of the exponent, the better
is the agreement with the discontinuous phase transition points. So, the
exponent can be an interesting indicator to locate continuous and
discontinuous phase transition points in surface reaction models. However,
further studies must be performed in this direction. From now on, we focus
only on the coefficient of determination. In the next subsection we present
our results for the protocol DIF-I and algorithm II.

\subsection{Algorithm II: Protocol DIF--I with \textquotedblleft
OR\textquotedblright\ adsorption process}

\label{subsec2}

\begin{figure*}[tbh]
\begin{center}
\includegraphics[width=0.83\columnwidth]{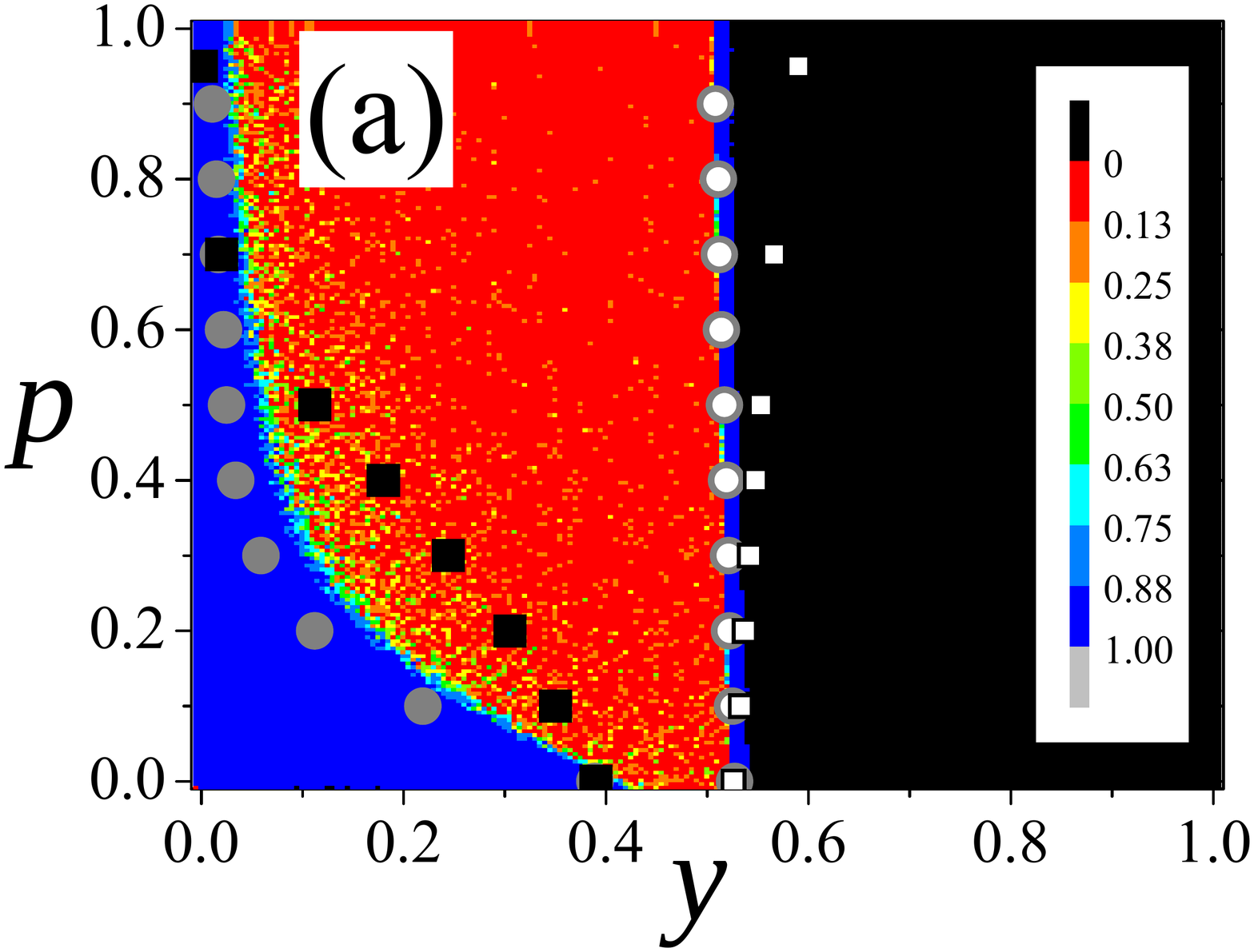} %
\includegraphics[width=0.83%
\columnwidth]{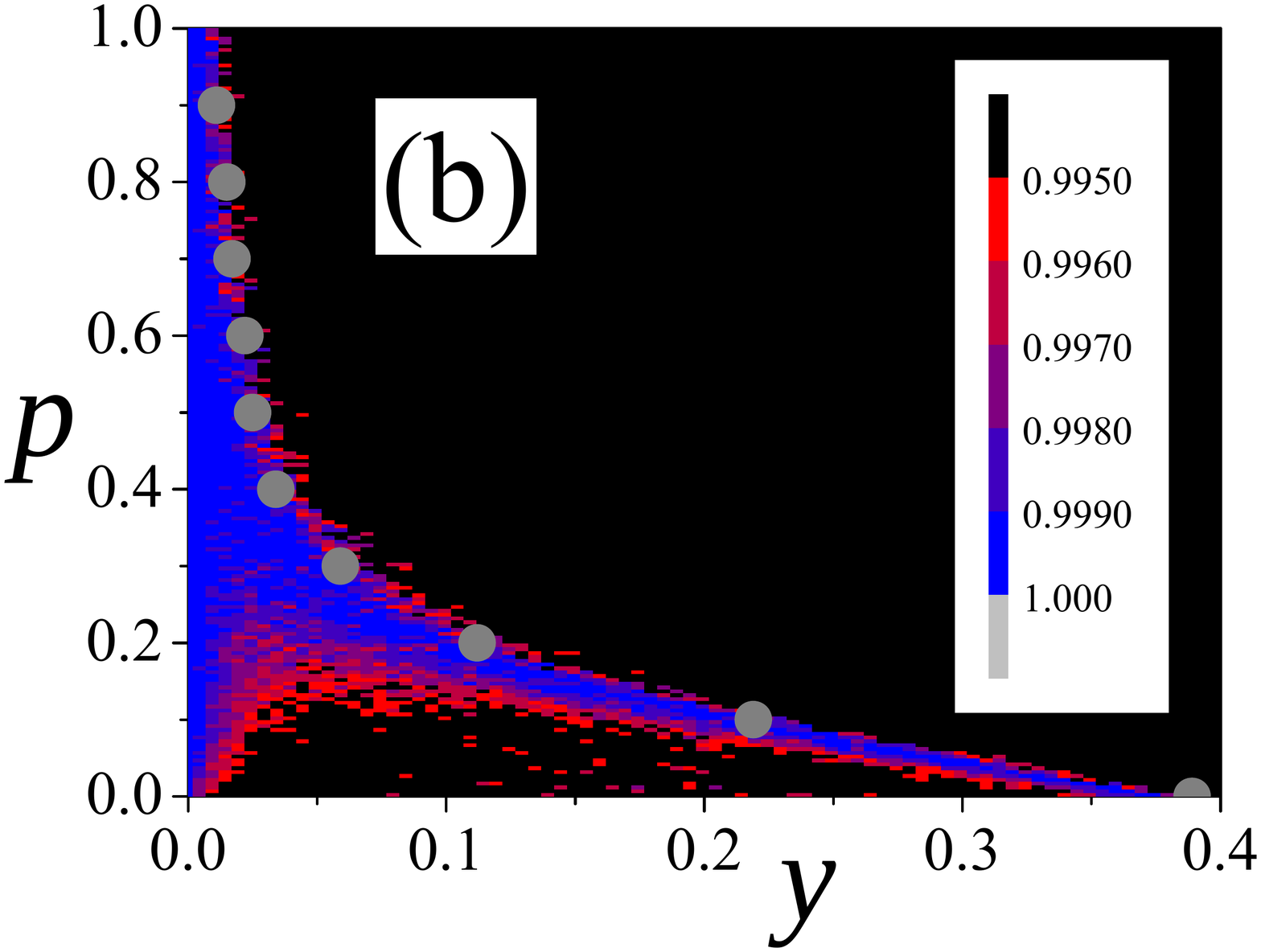}
\end{center}
\caption{(a) Coefficient of determination $r$ as function of $y$ and $p$
obtained with the protocol DIF-I and the algorithm II through TDMC
simulations, as well as the results of SSMC simulations. The filled and
empty black squares correspond, respectively, to the continuous and
discontinuous phase transitions predicted in Ref. \protect\cite{jensen1990a}%
, and the filled and empty filled gray circles correspond, respectively, to
the continuous and discontinuous transition points, both obtained via SSMC
simulations. (b) Refinement of the region of continuous phase transition
points.}
\label{Fig:Algorithm_2}
\end{figure*}

In this subsection, we perform simulations following the algorithm II in
which both diffusion and adsorption processes can occur in the same trial,
i.e., the diffusion does not exclude the possibility of the adsorption. As
can be seen in Fig. \ref{Fig:Algorithm_2}, the analysis of the results are
similar to those presented above, i.e., the TDMC simulations are also
supported by the SSMC simulations. We also show the estimates obtained in
Ref. \cite{jensen1990a} which are presented in Table \ref{Table:ssmc}.

The plots (a) and (b) of this figure correspond to the Fig. \ref{Fig:space1}
(a) and (b), respectively, and their format are similar. In fact, it is
important to notice that the discontinuous phase transition points of Fig. %
\ref{Fig:space1} remain practically the same as shown in Fig. \ref%
{Fig:Algorithm_2} (a). In this same plot, we show that the results of our
SSMC simulations is in complete agreement with those obtained via TDMC
simulations. Figure \ref{Fig:Algorithm_2} (b) presents a refinement of the
coefficient of determination for the region of the continuous phase
transition. As shown in Fig. \ref{Fig:space1}, this refinement narrows the
blue prolongation making clearer that the coefficient of determination
perfectly matches the gray points obtained by SSMC simulations. These gray
points are obtained exactly as those presented in Fig. \ref%
{Fig:density_by_SSMC} for the algorithm I, and therefore, we find it
unnecessary to show their figures here.

In addition, we also present the points obtained in Ref. \cite{jensen1990a}.
Those points do not match any of the regions of phase transition,
differently from what we find for algorithm I, in which its discontinuous
phase transition points are in complete agreement with ours. This difference
is expected since we are not dealing with the same algorithm. However, this
emphasizes an important point: the discontinuous line is more sensitive to
how the diffusion is combined with adsorption than the way the diffusion is
considered in the model.

\subsection{Algorithms III and IV, and the effects of $N_{run}$ and $L$ on
the coefficient of determination}

\label{subsec3}

Now, we combine the protocol DIF-II with the prescriptions I and II, with we
call algorithms III and IV, to study, as above, the influence of the
diffusion of the atoms/molecules on the lattice. Thus, following the last
two subsections, we obtain the coefficient of determination for the
algorithms III and IV, whose results are shown in Fig. \ref%
{Fig:Algs_III_and_IV}. The results obtained via SSMC simulations are also
present in this figure and in Table \ref{Table:ssmc}.

\begin{figure}[tbh]
\begin{center}
\includegraphics[width=0.5\columnwidth]{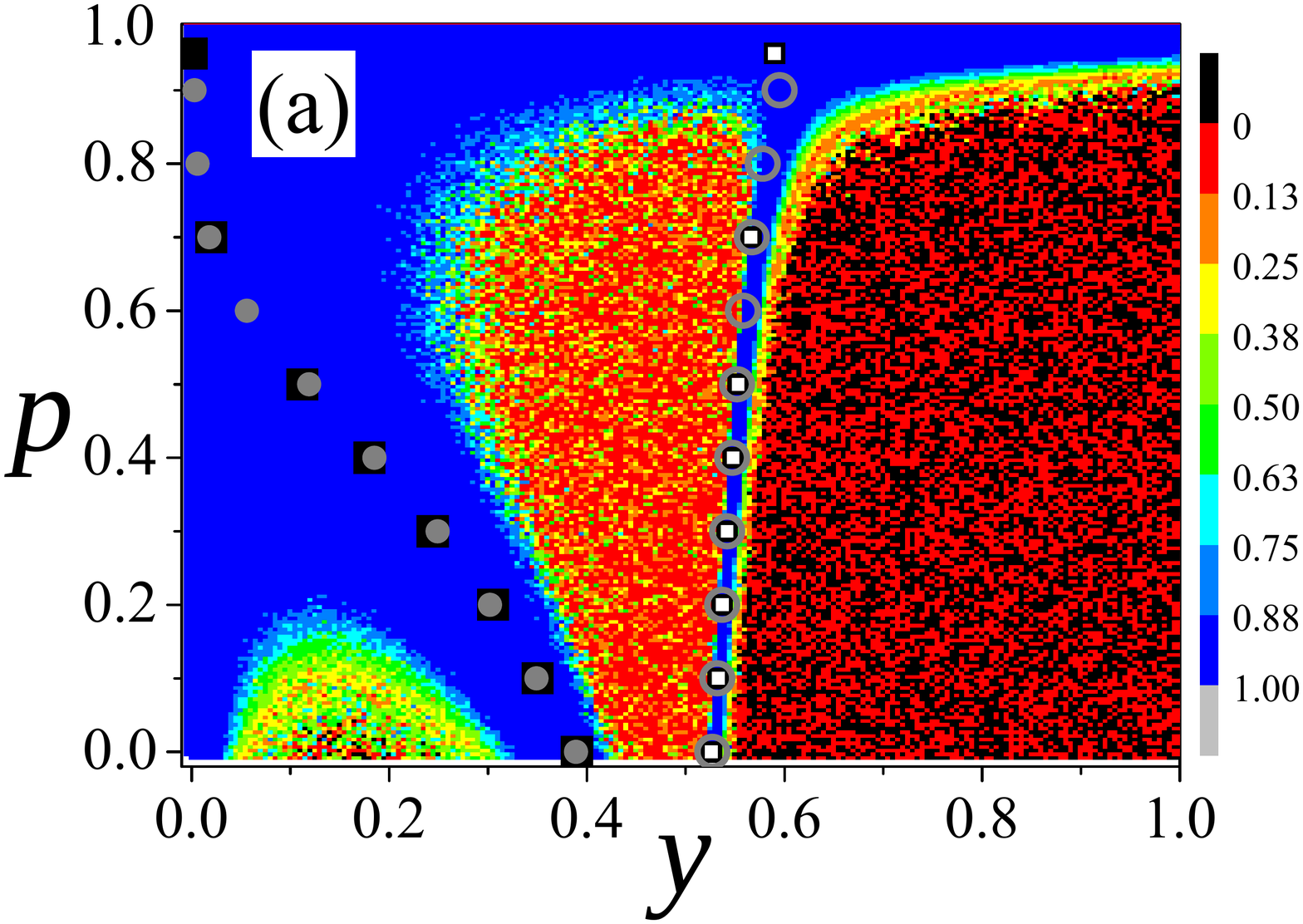}%
\includegraphics[width=0.5%
\columnwidth]{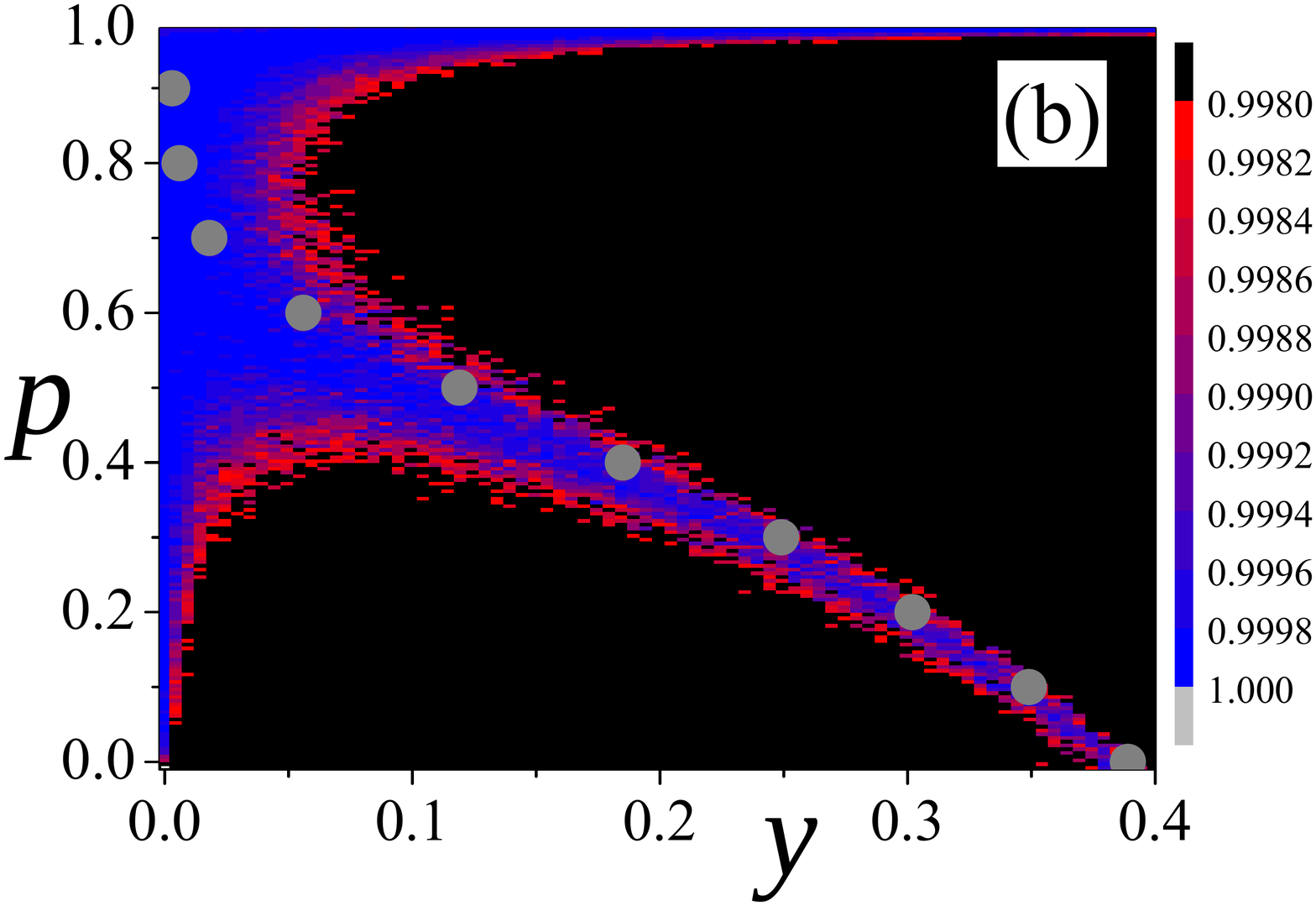} %
\includegraphics[width=0.5\columnwidth]{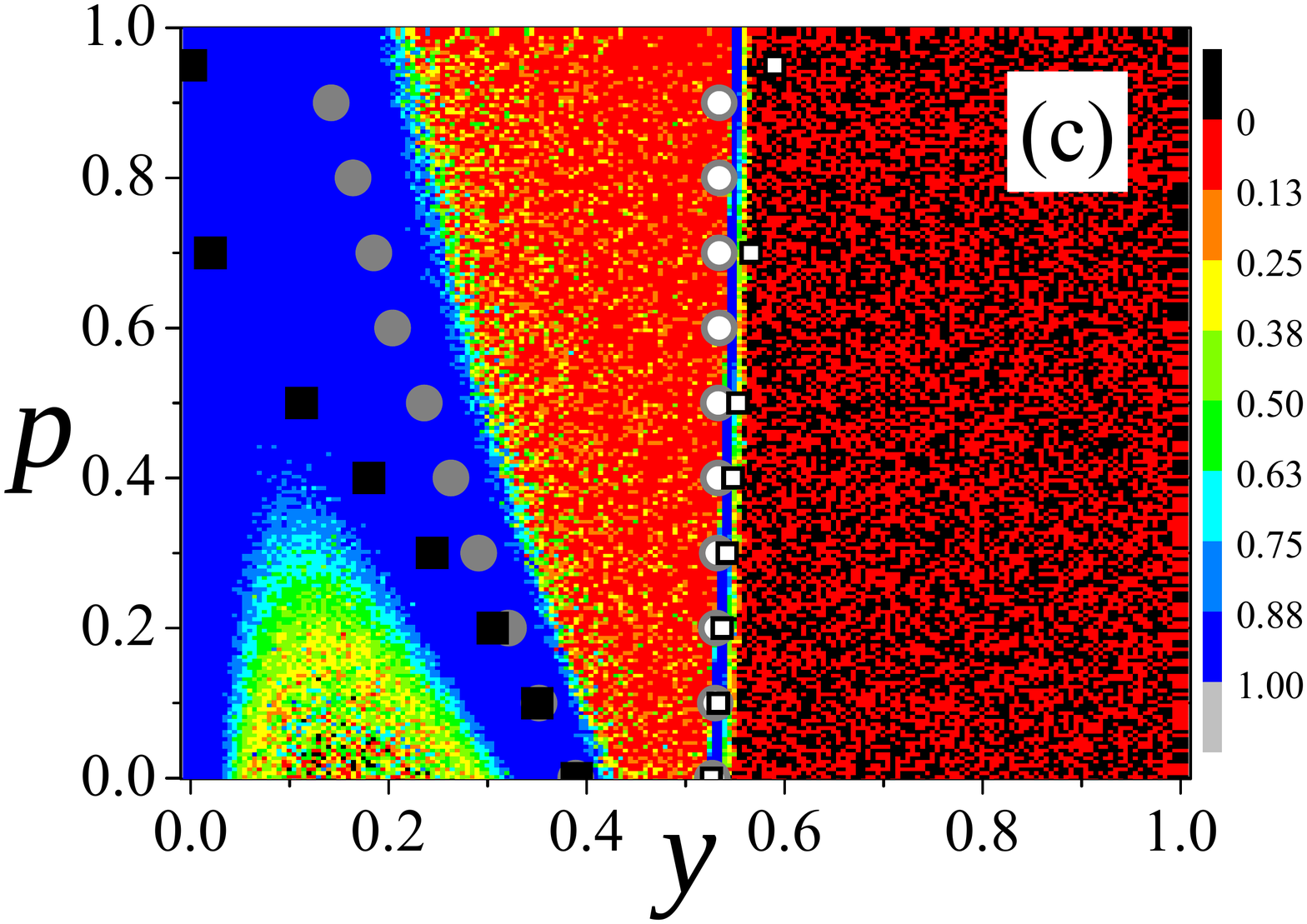}%
\includegraphics[width=0.5%
\columnwidth]{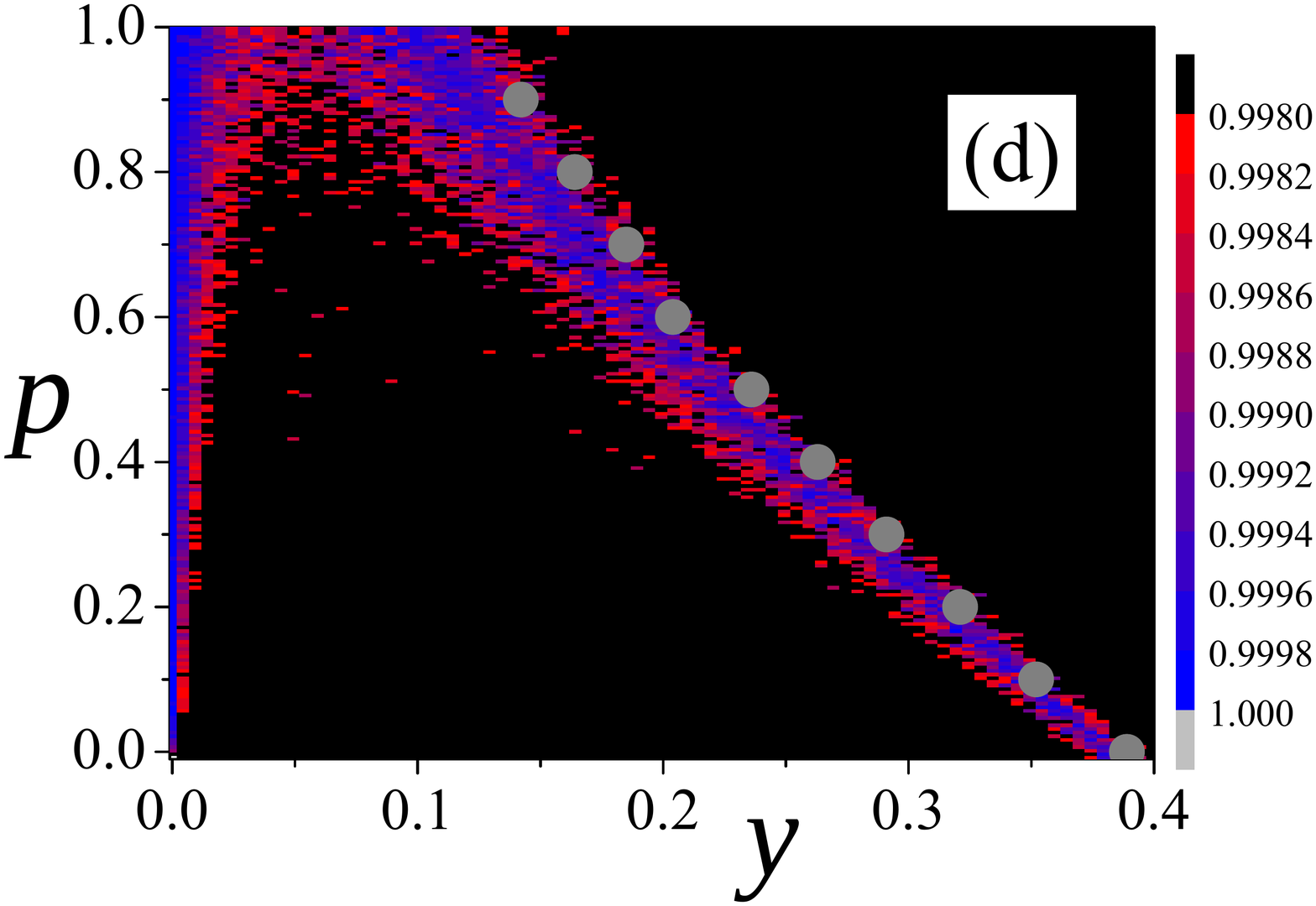}
\end{center}
\caption{Coefficient of determination $r$ as function of $y$ and $p$. The
algorithm III is represented by the plots (a) and (b), and the results for
the algorithm IV is shown in the plots (c) and (d). Following the same
procedure as in the previous figures, (a) and (c) are the complete diagrams
while (b) and (d) are the refinements of the regions corresponding to the
continuous phase transition.}
\label{Fig:Algs_III_and_IV}
\end{figure}

Figures \ref{Fig:Algs_III_and_IV} (a) and (b) represent the algorithm III
and correspond, respectively, to the complete diagram and the region of
continuous phase transition after the refinement procedure. From Fig. \ref%
{Fig:Algs_III_and_IV} (a), it is clear that the algorithm III must be the
corresponding model used in Ref. \cite{jensen1990a} since our points
obtained via SSMC simulations (grey circles) are in complete agreement with
the points extracted from that paper (black squares) in both regions. In
Fig. \ref{Fig:Algs_III_and_IV} (c) and (d), we show our results based on the
algorithm IV. In that case, we can observe that our estimates for the region
of continuous phase transition is completely different from those presented
in Ref. \cite{jensen1990a}. On the other hand, there is only a slightly
difference between them in the region of the discontinuous phase transition.

So, with the results of the four algorithms in hand, we can consider some
important points. First, the choice of XOR or OR to combine the diffusion
and adsorption processes seems to be more important to the region of
continuous phase transition that to the region of discontinuous one, since
we have a complete agreement between algorithms I and III which share the
same prescription (diffusion process XOR adsorption process) but possess
different protocols for the diffusion (DIF-I and DIF-II, respectively). In
addition, we have a certain similarity between the algorithms II and IV
which also share the same prescription (diffusion process OR adsorption
process) with different protocols, DIF-I\ and DIF-II, respectively. On the
other hand, the choice of the diffusion protocol leads to more similarity in
the region of continuous phase transition as can be observed between the
algorithms I and II and also between the algorithms III and IV.

For high diffusion rates, $p\rightarrow 1$, the rule XOR (prescription I)
completely suppress the possibility of the adsorption of atoms/molecules on
the lattice, generating false positive regions that makes it difficult to
locate the continuous phase transition points. So, only with the high
resolution obtained through the refinement process, we are able to locate
such points independently of what diffusion process is used.

\begin{figure}[tbh]
\begin{center}
\includegraphics[width=0.8\columnwidth]{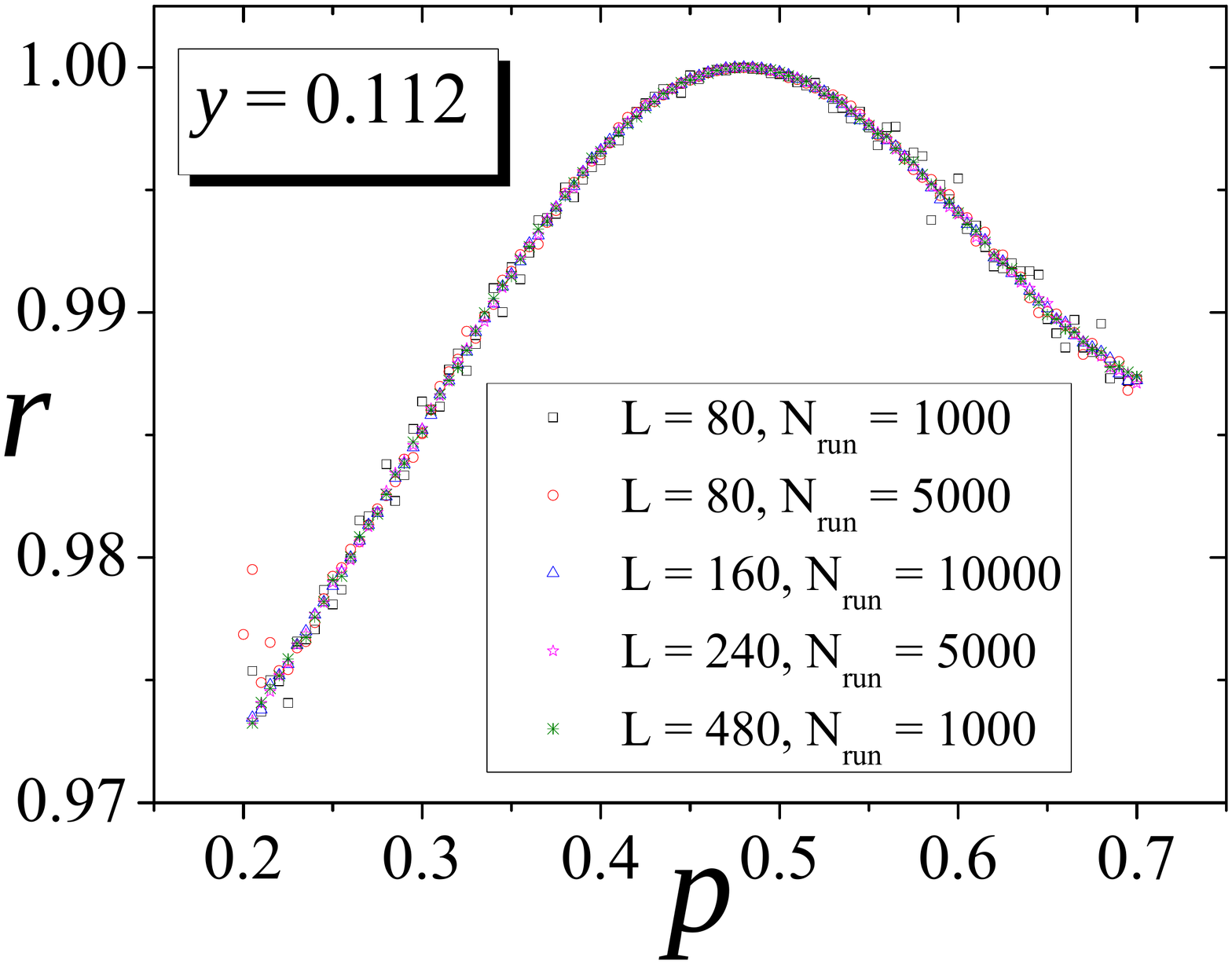}
\end{center}
\caption{Effects of $L$ and $N_{run}$ on the coefficient of determination $r$%
. The optimization curve $r\times p$ shows that there is no noticeable
difference among the five curves obtained for diferent values of $L$ and of $%
N_{run}$.}
\label{Fig:Nruns_and_L}
\end{figure}

Finally, important questions can be raised about the parameters considered
in our TDMC simulations such as the linear size of the square lattice, $L=80$%
, and the number of runs, $N_{run}=5000$. From these two parameters, the
lattice size seems to be the most relevant one since in Ref. \cite%
{jensen1990a} the authors consider lattices of linear size $L=500$ in the
study of the continuous phase transitions of the model. To supress any
doubts, we prepare as an example, a simple plot in which we fix the
adsorption rate in $y=0.112$ and calculate the coefficient of determination $%
r$ as function of $p$ for the algorithm III with several values of $L$ and $%
N_{run}$. According to the Ref. \cite{jensen1990a} is expected that the
maximun value of $r$, for sufficienly large lattices, should occur for $%
p\approx 0.5$. Figure \ref{Fig:Nruns_and_L} shows the curve $r$ $\times p$
for five different lattice sizes and three different number of runs. 
As can be observed, there is no significant difference among the curves,
mainly in the region of highest value of $r$. This example, along with other
simulations performed initially, shows that our choice of the parameters $%
L=80$ and $N_{run}=5000$ is appropriate and garantee that our results are
reliable and robust.

\section{Conclusions}

\label{sec:conclusions} In this work, we studied a modified version of the
Ziff-Gulari-Barshad (ZGB) model to include the spatial diffusion of the $O$
atoms and $CO$ molecules adsorbed on the surface through four different
algorithms. By using an alternative method that optimizes the coefficient of
determination, we were able to localize the regions of continuous and
discontinuous phase transitions of the model. Our results showed that, for
all algorithms, the critical behaviour is strongly changed by introducing
the diffusion of the atoms/molecules. On the contrary, the discontinuous
phase transition in much less sensitive to the diffusion even when the
diffusion rate is high. Our results were obtained through time-dependent
Monte Carlo (TDMC) simulations based on the optimization of the coefficient
of determination which took into account the time evolution of the two order
parameters of the model: the densities of vacant sites, $\rho_V$, and of $CO$
molecules, $\rho_{CO}$. All the results were corroborated by steady-state
Monte Carlo (SSMC) simulations.

The four algorithms were constructed in such a way that the diffusion and
adsorption processes are combined in two ways. In addition, there are also
two ways to implement the diffusion. We presented these algorithms as
algorithm I, II, III, and IV, and we showed that only the algorithm III
corresponds to the results found in the literature. In that case, only one
process is allowed in each trial, the diffusion process or the adsorption
process, and the diffusion of an atom/molecule changes with the number of
the empty sites in its neighbourhood.

In summary, this work classified the problem of diffusion in surface
reaction models (more precisely in the ZGB model) and showed that the
problem can be solved via TDMC simulations based on the optimization of a
simple statistical concept, the coefficient of determination. As show in the
figures, the TDMC simulations were in line with our SSMC simulations which
confirmed the efficiency of the coefficient of determination in estimating
the phase transitions of the model.

\section*{Acknowledgments}

This research work was in part supported financially by CNPq (National
Council for Scientific and Technological Development).

\end{document}